\documentclass{article}
\pdfoutput=1
\usepackage{a4wide,amsmath,graphics}
\newcommand{\F}[4]{\,_{#1}F_{#2}\left(\left.\begin{array}{c}#3\end{array}\right|#4\right)}
\DeclareMathOperator{\Tr}{Tr}

\title{Heavy quark form factors in the large $\beta_0$ limit}

\begin{document}

\newlength{\prepwidth}
\settowidth{\prepwidth}{MITP/17-026}
\begin{flushright}
\begin{minipage}{\prepwidth}
TTP17-022\\
MITP/17-026
\end{minipage}
\end{flushright}
\vspace{3mm}
\begin{center}
{\LARGE Heavy quark form factors in the large $\beta_0$ limit}\\[3mm]
A.\,G.~Grozin\footnote{A.G.Grozin@inp.nsk.su}\\
Budker Institute of Nuclear Physics, Novosibirsk, Russia%
\footnote{Also Novosibirsk State University, Novosibirsk, Russia};\\
Institut f\"ur Theoretische Teilchenphysik,\\
Karlsruher Institut f\"ur Technologie, Germany;\\
THEP, Institut f\"ur Physik,
Universit\"at Mainz,
Germany
\end{center}

\begin{abstract}
Heavy quark form factors are calculated at $\beta_0 \alpha_s \sim 1$
to all orders in $\alpha_s$ at the first order in $1/\beta_0$.
The $n_f^2 \alpha_s^3$ terms in the recent results~\cite{HSSS:17}
for the vector form factors are confirmed,
and $n_f^{L-1} \alpha_s^L$ terms for higher $L$ are predicted.
\end{abstract}

\section{Introduction}
\label{S:0}

Quark form factors are building blocks for various production cross sections and decay widths in QCD.
Recently massive-quark vector form factors have been calculated to to 3 loops~\cite{HSSS:17}.

\begin{figure}[ht]
\begin{center}
\includegraphics{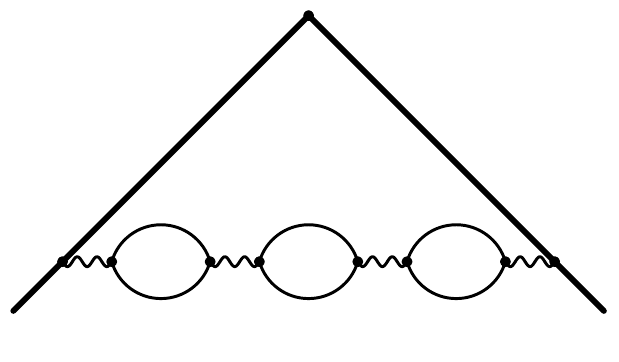}
\end{center}
\caption{Diagrams producing the highest degree of $n_f$ in each order of perturbation theory.}
\label{F:d}
\end{figure}

We'll consider heavy-quark form factors in the large $\beta_0$ limit,
where $\beta_0 \alpha_s \sim 1$, and $1/\beta_0$ is an expansion parameter
(see the reviews~\cite{B:99,G:04}).
A bare form factor can be written as
\begin{equation}
F = 1 + \sum_{L=1}^\infty \sum_{n=0}^{L-1} a_{Ln} \beta_0^n \left(\frac{g_0^2}{(4\pi)^{d/2}}\right)^L\,.
\label{0:F}
\end{equation}
Keeping terms with the highest degree of $\beta_0$ in each order of perturbation theory,
we get
\begin{equation}
F = 1 + \frac{1}{\beta_0} f\left(\frac{\beta_0 g_0^2}{(4\pi)^{d/2}}\right)
+ \mathcal{O}\left(\frac{1}{\beta_0^2}\right)\,.
\label{0:b}
\end{equation}
The leading coefficients $a_{L,L-1}$ can be easily obtained from $n_f^{L-1}$ terms
(Fig.~\ref{F:d}).
We shall consider only the first $1/\beta_0$ order%
\footnote{In some cases it is possible to obtain results for $1/\beta_0^2$ corrections,
see, e.\,g., \cite{B:93,G:16}.}.

\section{Heavy-quark bilinear currents}
\label{S:J}

We consider the QCD currents
\begin{equation}
J_0 = \bar{Q}_0 \Gamma Q_0 = Z(\alpha_s^{(n_f)}(\mu)) J(\mu)\,,\quad
\Gamma = \gamma^{[\mu_1} \cdots \gamma^{\mu_n]}\,,
\label{J:QCD}
\end{equation}
where $Q_0$ is a bare heavy-quark field.
The antisymmetrized product of $n$ $\gamma$ matrices has the property
\begin{equation}
\gamma^\mu \Gamma \gamma_\mu = \eta (d - 2 n) \Gamma\,,\quad
\eta = (-1)^n\,.
\label{J:h}
\end{equation}
In situations when the initial heavy-quark momentum $p_1$ and the final one $p_2$
can be written as $p_{1,2} = m v_{1,2} + k_{1,2}$
($m$ is the on-shell mass)
with $k_{1,2} \ll m$,
these currents can be expanded in HQET ones~\cite{FG:90,N:92}:
\begin{equation}
J(\mu) = \sum_{i=0}^2 H_i(\mu,\mu') \tilde{J}_i(\mu')
+ \frac{1}{2 m} \sum_i G_i(\mu,\mu') \tilde{O}_i(\mu')
+ \mathcal{O}\left(\frac{1}{m^2}\right)\,,
\label{J:HQET}
\end{equation}
where the leading HQET currents are
\begin{equation}
\tilde{J}_{i0} = \bar{h}_{v_2 0} \Gamma_i h_{v_1 0}
= \tilde{Z}(\alpha_s^{(n_l)}(\mu)) \tilde{J}_i(\mu)\,,\quad
\Gamma_i = \Gamma\,,\;\rlap/v_1 \Gamma + \Gamma \rlap/v_2\,,\;
\rlap/v_1 \Gamma \rlap/v_2\,,
\label{J:i}
\end{equation}
and $\tilde{O}_i$ are local and bilocal dimension-4 HQET operators
with appropriate quantum numbers.
The HQET current renormalization constant $\tilde{Z}$
does not depend on the Dirac structure
and is a function of the Minkowski angle $\vartheta$:
$v_1 \cdot v_2 = \cosh\vartheta = w$.

The coefficients in~(\ref{J:HQET}) can be obtained by matching
the on-shell matrix elements ($k_{1,2}=0$) in QCD and HQET:
\begin{equation}
\begin{split}
&{<}Q(p_2=mv_2)|J_0|Q(p_1=mv_1){>} = \sum_{i=0}^2 F_i\,\bar{u}_2 \Gamma_i u_1\,,\\
&{<}Q(k_2=0)|\tilde{J}_{i0}|Q(k_1=0){>} = \tilde{F}_i\,\bar{u}_2 \Gamma_i u_1\,,\quad
\tilde{F}_i =1
\end{split}
\label{J:F}
\end{equation}
(all loop corrections to $\tilde{F}_i$ vanish because they contain no scale).
Therefore the bare matching coefficients
(in the relation similar to~(\ref{J:HQET}) but for the bare currents)
are $H^0_i = F_i/\tilde{F}_i = F_i$.
The renormalized matching coefficients are
\begin{equation}
H_i(\mu,\mu') = H^0_i
\frac{\tilde{Z}(\alpha_s^{(n_l)}(\mu'))}{Z(\alpha_s^{(n_f)}(\mu))}
= \frac{F_i \tilde{Z}}{\tilde{F}_i Z}\,.
\label{J:H}
\end{equation}
UV divergences cancel in the ratio $F_i/Z$
as well as in the ratio $\tilde{F}_i/\tilde{Z}$.
Both $F_i$ and $\tilde{F}_i$ contain IR divergences which cancel in the ratio $F_i/\tilde{F}_i$
because HQET is constructed to reproduce the IR behaviour of QCD
($\tilde{F}_i$ have no loop corrections because their UV and IR divergences cancel each other).

The dependence of $H_i(\mu,\mu')$ on $\mu$ and $\mu'$ is determined by the RG equations.
Their solution can be written as
\begin{equation}
\begin{split}
H_i(\mu,\mu') ={}& \hat{H}_i
\left(\frac{\alpha_s^{(n_f)}(\mu)}{\alpha_s^{(n_f)}(\mu_0)}\right)^{\gamma_{n0}/(2\beta_0^{(n_f)})}
K_{\gamma_n}^{(n_f)}(\alpha_s^{(n_f)}(\mu))\\
&{}\times\left(\frac{\alpha_s^{(n_l)}(\mu')}{\alpha_s^{(n_l)}(\mu_0)}\right)^{-\tilde{\gamma}_0/(2\beta_0^{(n_l)})}
K_{-\tilde{\gamma}}^{(n_l)}(\alpha_s^{(n_l)}(\mu'))\,,
\end{split}
\label{J:RG}
\end{equation}
where for any anomalous dimension
$\gamma(\alpha_s) = \gamma_0 \alpha_s/(4\pi) + \gamma_1 (\alpha_s/(4\pi))^2 + \cdots$
we define
\begin{equation}
K_\gamma(\alpha_s)
= \exp \int_0^{\alpha_s} \frac{d\alpha_s}{\alpha_s}
\left( \frac{\gamma(\alpha_s)}{2\beta(\alpha_s)} - \frac{\gamma_0}{2\beta_0} \right)
= 1 + \frac{\gamma_0}{2\beta_0}
\left( \frac{\gamma_1}{\gamma_0} - \frac{\beta_1}{\beta_0} \right)
\frac{\alpha_s}{4\pi} + \cdots
\label{J:K}
\end{equation}

Matrix elements of the currents with $n=0$, 1 can be written via smaller numbers of form factors:
\begin{equation}
{<}Q(mv_2)|J|Q(mv_1){>} = F^S \bar{u}_2 u_1\,,\quad
F^S = F_0 + 2 F_1 + (2w-1) F_2
\label{J:S}
\end{equation}
(where $F_i$ with $n=0$, $\eta=1$ are used), and
\begin{equation}
\begin{split}
&{<}Q(mv_2)|J^\mu|Q(mv_1){>} = (F^V_1 + F^V_2) \bar{u}_2 \gamma^\mu u_1 - F^V_2 \bar{u}_2 u_1 \frac{(v_1 + v_2)^\mu}{2}\,,\\
&F^V_1 = F_0 + 2 F_1 - (2w-3) F_2\,,\quad
F^V_2 = - 4 (F_1 + F_2)
\end{split}
\label{J:V}
\end{equation}
(where $F_i$ with $n=1$, $\eta=-1$ are used).

\section{Inversion relations}
\label{S:I}

\begin{figure}[ht]
\begin{center}
\begin{picture}(94,27)
\put(21,16){\makebox(0,0){\includegraphics{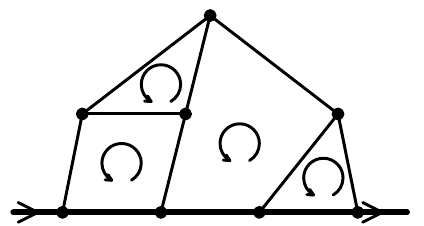}}}
\put(73,16){\makebox(0,0){\includegraphics{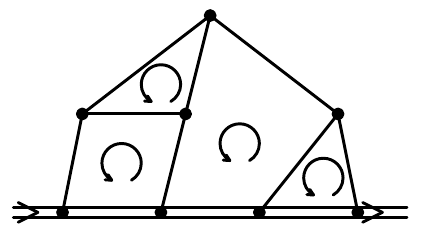}}}
\put(21,0){\makebox(0,0)[b]{a}}
\put(3.5,3){\makebox(0,0){$mv$}}
\put(38.5,3){\makebox(0,0){$mv$}}
\put(12,11){\makebox(0,0){$k_1$}}
\put(16,19){\makebox(0,0){$k_2$}}
\put(24,13){\makebox(0,0){$k_3$}}
\put(32.5,9.5){\makebox(0,0){$k_4$}}
\put(55.5,3){\makebox(0,0){$\omega$}}
\put(90.5,3){\makebox(0,0){$\omega$}}
\put(64,11){\makebox(0,0){$k_1$}}
\put(68,19){\makebox(0,0){$k_2$}}
\put(76,13){\makebox(0,0){$k_3$}}
\put(84.5,9.5){\makebox(0,0){$k_4$}}
\put(73,0){\makebox(0,0)[b]{b}}
\end{picture}
\end{center}
\caption{On-shell massive self-energy integrals and off-shell HQET ones.}
\label{F:Is}
\end{figure}
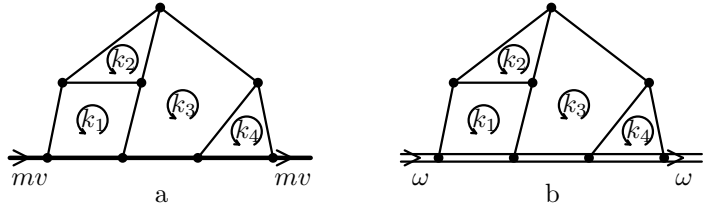

\begin{figure}[ht]
\begin{center}
\begin{picture}(54,12)
\put(13,6){\makebox(0,0){\includegraphics{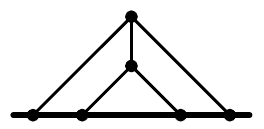}}}
\put(41,6){\makebox(0,0){\includegraphics{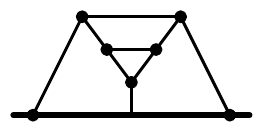}}}
\end{picture}
\end{center}
\caption{Examples of on-shell massive diagrams which cannot be transformed to off-shell HQET ones
by inversion relations.}
\label{F:Bad}
\end{figure}

On-shell massive self-energy integrals with one massive line
and any number of massless ones in some cases
can be expressed via similar off-shell HQET integrals.
Suppose all massless lines can be drawn at one side of the massive one
and the resulting graph is planar (e.\,g., the diagram in Fig.~\ref{F:Is}a).
Lines of such a diagram subdivide the plane into a number of polygonal cells
(plus the exterior);
with each cell we can associate a loop momentum (flowing counterclockwise).
Then outer massless edges of the diagram correspond to the denominators
$- k_i^2 - i0$; inner massless edges -- to $- (k_i-k_j)^2 - i0$;
and massive edges -- to $m^2 - (k_i + mv)^2 - i0$ (Table~\ref{Tab}).
The corresponding HQET diagram (Fig.~\ref{F:Is}b) has HQET denominators
$- 2 k_i\cdot v - 2 \omega - i0$ instead of massive ones.
First we perform Wick rotation of all loop momenta $k_{i0} \to i k_{i0}$
(in the $v$ rest frame).
Then, in Euclidean momentum space, we invert each loop momentum~\cite{BG:95a}:
\begin{equation}
k_i \to \frac{k_i}{k_i^2}\,.
\label{I:inv}
\end{equation}
Inversion transforms massive denominators to HQET ones (and vice versa)
if we identify
\begin{equation}
- 2 \omega = m^{-1}\,,
\label{I:omega}
\end{equation}
see Table~\ref{Tab}.
As a result, a massive on-shell diagram (Fig.~\ref{F:Is}a)
becomes $m^{-\sum n_i}$ (the sum runs over all massive line segments,
$n_i$ are their indices, i.\,e.\ the powers of the denominators)
times the off-shell HQET diagram (Fig.~\ref{F:Is}b)
with $\omega = - (2m)^{-1}$~(\ref{I:omega}).
The indices of all inner massless edges,
as well as of all massive edges (which become HQET ones),
remain intact (see Table~\ref{Tab}).
From the same table it is clear that
the index of an outer massless edge becomes $d - \sum n_i$,
where the sum runs over all edges of the cell to which this outer edge belongs
(they can be all massless, or one of them can be massive).
If there is a cell $k_i$ bounded only by inner massless edges,
and maybe one massive one,
then the denominator $(k_i^2)^{d-\sum n_j}$ will appear (Fig.~\ref{F:Bad}).
This denominator does not correspond to any line,
and hence the resulting integral is not a Feynman integral at all;
in this case, the discussed relation becomes rather useless (though formally correct).
The inversion relations~\cite{BG:95a} were used, e.\,g., in~\cite{CM:02,G:04a}).

\begin{table}[h]
\caption{Inversion relations.}
\label{Tab}
\begin{center}
\begin{tabular}{|c|c|c|c|}
\hline
& Minkowski & Euclidean & Inversion\\
\hline
outer massless & $- k_i^2 - i0$ & $k_i^2$ & $\displaystyle \frac{1}{k_i^2}$\\
\hline
inner massless & $- (k_i-k_j)^2 -i0$ & $(k_i-k_j)^2$ & $\displaystyle \frac{(k_i-k_j)^2}{k_i^2 k_j^2}$\\
\hline
massive & $- k_i^2 - 2 m v\cdot k_i - i0$ & $k_i^2 - 2 i m k_{i0}$ & $\displaystyle m \frac{- 2 \omega - 2 i k_{i0}}{k_i^2}$\\
\hline
HQET & $- 2 \omega - 2 k_i\cdot v -i0$ & $- 2 \omega - 2 i k_{i0}$ & $\displaystyle m^{-1} \frac{k_i^2 - 2 i k_{i0}}{k_i^2}$\\
\hline
measure & $d^d k_i$ & $i d^d k$ & $\displaystyle i \frac{d^d k_i}{(k_i^2)^d}$\\
\hline
\end{tabular}
\end{center}
\end{table}

\begin{figure}[ht]
\begin{center}
\begin{picture}(130,22)
\put(30,11){\makebox(0,0){\includegraphics{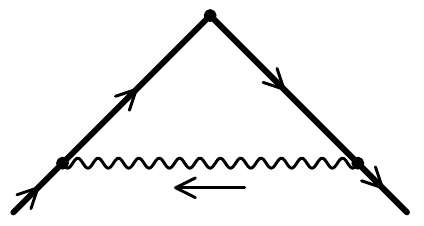}}}
\put(6,3.5){\makebox(0,0){$m_1 v_1$}}
\put(54,3.5){\makebox(0,0){$m_2 v_2$}}
\put(12,13.5){\makebox(0,0){$k+m_1 v_1$}}
\put(48,13.5){\makebox(0,0){$k+m_2 v_2$}}
\put(30,1){\makebox(0,0){$k$}}
\put(100,11){\makebox(0,0){\includegraphics{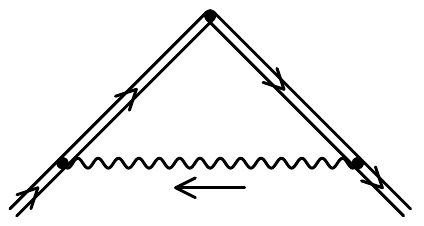}}}
\put(77,3.5){\makebox(0,0){$\omega_1$}}
\put(123,3.5){\makebox(0,0){$\omega_2$}}
\put(82,13.5){\makebox(0,0){$k\cdot v_1+\omega_1$}}
\put(118,13.5){\makebox(0,0){$k\cdot v_2+\omega_2$}}
\put(100,1){\makebox(0,0){$k$}}
\end{picture}
\end{center}
\caption{One-loop vertex integrals.}
\label{F:I1}
\end{figure}
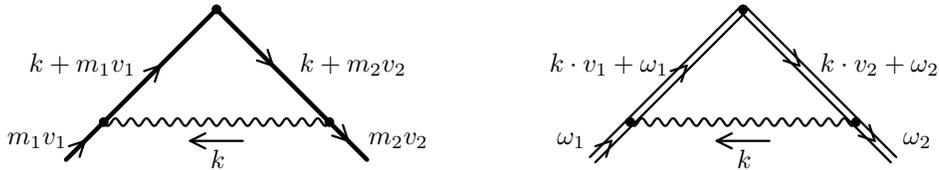

The inversion relations can be generalized to similar vertex integrals;
the masses of the initial particle and the final one may differ.
At one loop (Fig.~\ref{F:I1}), the integrals
\begin{align}
&M(n_1,n_2,n;\vartheta;m_1,m_2) =
\int \frac{d^d k}{i\pi^{d/2}}
\nonumber\\
&\quad{}\times\frac{1}%
{[- k^2 - 2 m_1 v_1 \cdot k - i0]^{n_1} [- k^2 - 2 m_2 v_2 \cdot k - i0]^{n_2} (-k^2-i0)^n}\,,
\label{I:M}\\
&I(n_1,n_2,n;\vartheta;\omega_1,\omega_2) =
\int \frac{d^d k}{i\pi^{d/2}}
\nonumber\\
&\quad{}\times\frac{1}%
{[- 2 k \cdot v_1 - 2 \omega_1 - i0]^{n_1} [- 2 k \cdot v_2 - 2 \omega_2 - i0]^{n_2} (-k^2-i0)^n}
\label{I:I}
\end{align}
are related by
\begin{equation}
M(n_1,n_2,n;\vartheta;m_1,m_2) = m_1^{-n_1} m_2^{-n_2}
I(n_1,n_2,d-n_1-n_2-n;\vartheta; - (2 m_1)^{-1},- (2 m_2)^{-1})\,.
\label{I:R}
\end{equation}

The integrals $I$~(\ref{I:I}) have been investigated in~\cite{GK:11}.
Here we need only the integrals $M$~(\ref{I:M}) with $m_1=m_2$;
they reduce to the integrals $I$~(\ref{I:I}) with $\omega_1=\omega_2$
which are especially simple~\cite{GK:11}:
\begin{equation}
I(n_1,n_2,n;\vartheta;\omega,\omega) = (-2\omega)^{d-n_1-n_2-2n} I(n_1+n_2,n)
\F{3}{2}{n_1,n_2,\frac{d}{2}-n\\\frac{n_1+n_2}{2},\frac{n_1+n_2+1}{2}}{\frac{1-\cosh\vartheta}{2}}\,,
\label{I:F}
\end{equation}
where
\begin{equation}
I(n_1,n) = \frac{\Gamma(-d+n_1+2n) \Gamma(d/2-n)}{\Gamma(n_1) \Gamma(n)}
\label{I:1}
\end{equation}
is the one-loop HQET self-energy integral.
We only need integer $n_{1,2}$;
in this case all $I$ reduce by IBP to 2 master integrals~\cite{GK:11}:
$I(1,0,n)$ (trivial) and $I(1,1,n)$ (given by~(\ref{I:F})).

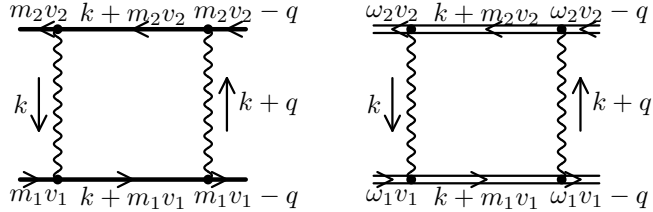
\begin{figure}[h]
\begin{center}
\begin{picture}(85,28)
\put(19,14){\makebox(0,0){\includegraphics{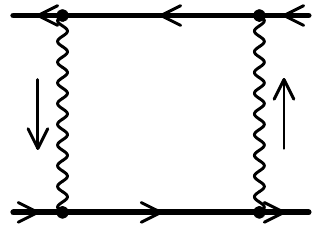}}}
\put(6.5,1.5){\makebox(0,0){$m_1 v_1$}}
\put(6.5,26){\makebox(0,0){$m_2 v_2$}}
\put(34,1.5){\makebox(0,0){$m_1 v_1 - q$}}
\put(34,26){\makebox(0,0){$m_2 v_2 - q$}}
\put(19,1.5){\makebox(0,0){$k + m_1 v_1$}}
\put(19,26){\makebox(0,0){$k + m_2 v_2$}}
\put(5,14){\makebox(0,0)[r]{$k$}}
\put(33,14){\makebox(0,0)[l]{$k+q$}}
\put(66,14){\makebox(0,0){\includegraphics{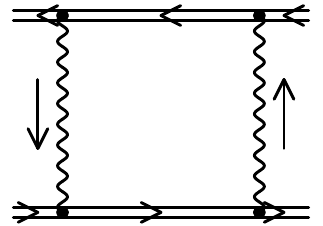}}}
\put(53.5,1.5){\makebox(0,0){$\omega_1 v_1$}}
\put(53.5,26){\makebox(0,0){$\omega_2 v_2$}}
\put(81,1.5){\makebox(0,0){$\omega_1 v_1 - q$}}
\put(81,26){\makebox(0,0){$\omega_2 v_2 - q$}}
\put(66,1.5){\makebox(0,0){$k + m_1 v_1$}}
\put(66,26){\makebox(0,0){$k + m_2 v_2$}}
\put(52,14){\makebox(0,0)[r]{$k$}}
\put(80,14){\makebox(0,0)[l]{$k+q$}}
\end{picture}
\end{center}
\caption{Box diagrams.}
\label{F:B}
\end{figure}

Inversion relations can be generalized to diagrams with more external legs.
For example, the one-loop massive box diagram with 2 on-shell legs
and the corresponding off-shell HQET one (Fig.~\ref{F:B})
\begin{align}
&M(n_1,n_2,n_3,n_4;\vartheta;m_1,m_2;q^2,q\cdot v_1,q\cdot v_2) = \int \frac{d^d k}{i\pi^{d/2}} \times{}
\nonumber\\
&\frac{1}%
{(- k^2 - 2 m_1 v_1\cdot k)^{n_1} (- k^2 - 2 m_2 v_2\cdot k)^{n_2} (- (k+q)^2)^{n_3} (- k^2)^{n_4}}\,,
\label{I:BM}\\
&I(n_1,n_2,n_3,n_4;\vartheta;\omega_1,\omega_2;q^2,q\cdot v_1,q\cdot v_2) = \int \frac{d^d k}{i\pi^{d/2}} \times{}
\nonumber\\
&\frac{1}%
{(- 2 k\cdot v_1 - 2 \omega_1)^{n_1} (- 2 k\cdot v_2)^{n_2} (- (k+q)^2)^{n_3} (- k^2)^{n_4}}
\label{I:BI}
\end{align}
are related by
\begin{align}
&M(n_1,n_2,n_3,n_4;\vartheta;m_1,m_2;q^2,q\cdot v_1,q\cdot v_2)
= m_1^{-n_1} m_2^{-n_2} (-q^2)^{n_3}
\label{I:BR}\\
&I(n_1,n_2,n_3,d-n_1-n_2-n_3-n_4;\vartheta;
-(2 m_1)^{-1},-(2 m_2)^{-1};1/q^2,q\cdot v_1/(-q^2),q\cdot v_2/(-q^2))\,.
\nonumber
\end{align}

\section{Large-$\beta_0$ limit}
\label{S:b}

We need only terms with the highest degree of $n_f$;
therefore, there is no need to distinguish between $n_f$ and $n_l=n_f-1$,
or any $n_f+\text{const}$.
The gluon propagator can be written as
\begin{equation}
D_{\mu\nu}(k) = \frac{1}{k^2 (1-\Pi(k^2))} \left(g_{\mu\nu} - \frac{k_\mu k_\nu}{k^2}\right)\,,
\label{b:G}
\end{equation}
where the gluon self energy is
\begin{equation}
\begin{split}
\Pi(k^2) &{}= \beta_0 \frac{g_0^2}{(4\pi)^{d/2}} e^{-\gamma\varepsilon} \frac{D(\varepsilon)}{\varepsilon} (-k^2)^{-\varepsilon}\,,\\
D(\varepsilon) &{}= e^{\gamma\varepsilon}
\frac{(1-\varepsilon) \Gamma(1+\varepsilon) \Gamma^2(1-\varepsilon)}{(1-2\varepsilon) (1-\frac{2}{3}\varepsilon) \Gamma(1-2\varepsilon)}
= 1 + \frac{5}{3} \varepsilon + \cdots
\end{split}
\label{b:D}
\end{equation}
At this leading large $\beta_0$ order,
the coupling constant renormalization is simple:
\begin{equation}
\beta_0 \frac{g_0^2}{(4\pi)^{d/2}} e^{-\gamma\varepsilon} = b Z_\alpha(b) \mu^{2\varepsilon}\,,\quad
b = \beta_0 \frac{\alpha_s(\mu)}{4\pi}\,,\quad
Z_\alpha = \frac{1}{1+b/\varepsilon}\,.
\label{b:Za}
\end{equation}

The bare QCD matrix elements can be written in the form~\cite{PMP:84,B:93}
\begin{equation}
F_i = \delta_{i0} + \frac{1}{\beta_0} \sum_{L=1}^\infty \frac{f_i(\varepsilon,L\varepsilon)}{L} \Pi(-m^2)^L
+ \mathcal{O}\left(\frac{1}{\beta_0^2}\right)\,.
\label{b:f}
\end{equation}
It is convenient to write the functions $f_i(\varepsilon,u)$
in the form usual for on-shell massive QCD problems (see~\cite{G:04})
\begin{equation}
f_i(\varepsilon,u) = C_F \frac{e^{\gamma\varepsilon}}{D(\varepsilon)}
\frac{\Gamma(1-2u) \Gamma(1+u)}{\Gamma(3-u-\varepsilon)} N_i(\varepsilon,u)\,.
\label{b:fN}
\end{equation}

We calculate the vertex function (Fig.~\ref{F:d})
and multiply it by $Z_Q^{\text{os}}$ with the $1/\beta_0$ accuracy (see~\cite{G:04}).
Reducing on-shell massive QCD integrals to off-shell HQET ones by the inversion relation~(\ref{I:R})
and then to the master integrals by IBP~\cite{GK:11}, we obtain
\begin{align}
N_0(\varepsilon,u) ={}& \biggl[
- \eta u \frac{n-2+\varepsilon}{w-1}
- 2 (w+1) u (n-2)^2
- u \bigl( \eta u + 4 (w+1) \varepsilon \bigr) (n-2)
\biggr.
\nonumber\\
&\quad\biggl.{}
+ 2 (2-u) \bigl( w + (w+1) u \bigr)
- ( 6 w + 2 u + \eta u^2 ) \varepsilon
+ 2 \bigl( w - (w+1) u \bigr) \varepsilon^2
\biggr] F
\nonumber\\
&{} + \eta u \frac{n-2+\varepsilon}{w-1}
+ 2 (n-2)^2 + 4 \varepsilon (n-2)
- 6 (1-u^2) + 2 (1-u) (5+2u) \varepsilon - 2 (1-2u) \varepsilon^2\,,
\nonumber\\
N_1(\varepsilon,u) ={}& u \biggl[
\eta w \frac{n-2+\varepsilon}{w-1}
- \eta u (n-2) - 2 + u + \varepsilon - \eta u \varepsilon
\biggr] F
- \eta u \frac{n-2+\varepsilon}{w-1}\,,
\nonumber\\
N_2(\varepsilon,u) ={}& \eta u \frac{n-2+\varepsilon}{w-1}
\biggl[ 1 - \bigl( 1 + (w-1) u \bigr) F \biggr]\,,
\label{b:N}
\end{align}
where
\begin{equation}
F = \F{2}{1}{1,1+u\\3/2}{\frac{1-w}{2}}
\label{b:F}
\end{equation}
(the same function appears also in the 1-loop self-energy integral with arbitrary masses $m_{1,2}$ and arbitrary $p^2$,
where both indices are equal to 1~\cite{DK:01}).
At $\vartheta=0$ this result agrees with the result of~\cite{N:95} at $m_1=m_2$,
see also~\cite{G:04}%
\footnote{Note a typo: the unnumbered formula below~(8.93) should read
\[R_0 = \cosh(Lu)\,,\quad
R_1 = \frac{\sinh[(1-2u)L/2]}{\sinh(L/2)}\,.
\]}.

Re-expressing the bare form factors~(\ref{b:f}) via the renormalized coupling
we obtain
\begin{equation}
F_i = \delta_{i0} + \frac{1}{\beta_0} \sum_{L=1}^\infty \frac{f_i(\varepsilon,L\varepsilon)}{L}
\left[D(\varepsilon) \left(\frac{\mu^2}{m^2}\right)^\varepsilon \frac{b}{\varepsilon+b}\right]^L\,.
\label{b:fr}
\end{equation}
We should have (see~(\ref{J:H}))
\begin{equation}
\log F_0 = \log(Z(\alpha_s(\mu))/\tilde{Z}(\alpha_s(\mu))) + \log H(\mu,\mu)\,:
\label{b:r}
\end{equation}
negative degrees of $\varepsilon$ go to $\log(Z/\tilde{Z})$,
non-negative ones -- to $\log H$.
The function
\begin{equation}
f_0(\varepsilon,u) D(\varepsilon)^{u/\varepsilon} \left(\frac{\mu^2}{m^2}\right)^u
= \sum_{n,m=0}^\infty f_{nm} \varepsilon^n u^m
\label{b:feu}
\end{equation}
is regular at the origin;
expanding $(b/(\varepsilon+b))^L$ in $b$, we obtain a quadruple sum.
In the coefficient of $\varepsilon^{-1}$ all $f_{nm}$ except $f_{n0}$ cancel;
differentiating this coefficient in $\log b$
(and using the fact that $F$~(\ref{b:F}) at $u=0$ is $\vartheta/\sinh\vartheta$)
we obtain the anomalous dimension corresponding to $Z/\tilde{Z}$~\cite{PMP:84,B:93}:
\begin{equation}
\gamma_n - \tilde{\gamma} = - 2 \frac{b}{\beta_0} f_0(-b,0)
+ \mathcal{O}\left(\frac{1}{\beta_0^2}\right)\,.
\label{b:gamma}
\end{equation}
These anomalous dimensions at the $1/\beta_0$ order are~\cite{BG:95,BB:95}
\begin{align}
&\gamma_n = 4 C_F \frac{b}{\beta_0}
\frac{(1 + \frac{2}{3} b) \Gamma(2+2b)}{(1+b)^2 (2+b) \Gamma^3(1+b) \Gamma(1-b)}
(n-1) (3-n+2b)
+ \mathcal{O}\left(\frac{1}{\beta_0^2}\right)\,,
\label{b:gn}\\
&\tilde{\gamma} = 4 C_F \frac{b}{\beta_0}
\frac{(1 + \frac{2}{3} b) \Gamma(2+2b)}{(1+b) \Gamma^3(1+b) \Gamma(1-b)}
(\vartheta\coth\vartheta - 1)
+ \mathcal{O}\left(\frac{1}{\beta_0^2}\right)\,.
\label{b:gh}
\end{align}
Our results satisfy this requirement
($f_{1,2}(-b,0)=0$ because the QCD current $J$ does not mix with currents
with other Dirac structures).

In the coefficient of $\varepsilon^0$ all $f_{nm}$ except $f_{n0}$ and $f_{0m}$ cancel.
The coefficients $f_{n0}$ form $K_{\gamma_n-\tilde{\gamma}}(\alpha_s(\mu))$, see~(\ref{J:RG});
we have~\cite{B:93}
\begin{equation}
\hat{H}_i = \delta_{i0} + \frac{1}{\beta_0} \int_0^\infty du\,e^{-u/b} S_i(u)
+ \mathcal{O}\left(\frac{1}{\beta_0^2}\right)\,,
\label{b:hh}
\end{equation}
where the Borel images of the perturbative series for $\hat{H}_i$ are
\begin{equation}
S_i(u) = \frac{1}{u} \left[\left(e^{5/3} \frac{\mu_0^2}{m^2}\right)^u f_i(0,u) - f_i(0,0)\right]\,.
\label{b:S}
\end{equation}
The integral~(\ref{b:hh}) is not well defined because of poles at the integration contour.
The leading renormalon ambiguities are given by the residues at $u=1/2$~\cite{NS:95}
(see also~\cite{G:04}).
It is easy to calculate these residues because $F$~(\ref{b:F}) at $u=1/2$ is just $2/(w+1)$:
\begin{equation}
\Delta H_0 = \left(\frac{4}{w+1} - 3\right) \frac{\Delta\bar{\Lambda}}{2m}\,,\quad
\Delta H_1 = \frac{1}{w+1} \frac{\Delta\bar{\Lambda}}{2m}\,,\quad
\Delta H_2 = 0\,,
\label{b:amb}
\end{equation}
where
\begin{equation*}
\Delta\bar{\Lambda} = - 2 \frac{C_F}{\beta_0} e^{5/6} \Lambda_{\overline{\text{MS}}}\,.
\end{equation*}
As demonstrated in~\cite{NS:95},
these IR renormalon ambiguities are compensated by the UV renormalon ambiguities
in the matrix elements of the HQET operators $\tilde{O}_i$ in~(\ref{J:HQET}).

The hypergeometric function $F$~(\ref{b:F}) has been expanded in $u$ to all orders~\cite{DK:01},
the coefficients are expressed via Nielsen polylogarithms $S_{nm}(x)$.
The result~\cite{DK:01} is written for the case of an Euclidean angle%
\footnote{M.\,Yu.~Kalmykov has informed me that there is a typo:
the power of $\cos\vartheta$ in~(2.7) should be $1+2\varepsilon$.
This typo has been corrected in~\cite{DK:04}.};
its analytical continuation to Minkowski angles is
\begin{equation}
\begin{split}
F = \frac{1}{\sinh\vartheta (2 \cosh(\vartheta/2))^{2 u}}
\biggl[ \frac{\sinh(\vartheta u)}{u}
&{} - e^{-\vartheta u} \sum_{n=1}^\infty u^n \sum_{m=1}^n (-2)^{n-m} S_{m,n-m+1}(-e^{\vartheta})\\
&{} + e^{\vartheta u} \sum_{n=1}^\infty u^n \sum_{m=1}^n (-2)^{n-m} S_{m,n-m+1}(-e^{-\vartheta})
\biggr]\,.
\end{split}
\label{b:DK}
\end{equation}
It is possible to re-express this expansion in terms of Nielsen polylogarithms of just one argument,
see~\cite{DD:84},
but then the symmetry $\vartheta\to-\vartheta$ will not be explicit.

\textbf{Acknowledgements}.
I am grateful to M.~Steinhauser for useful comments
and hospitality in Karlsruhe, where the major part of this work was done;
to J.\,M.~Henn for useful discussions and hospitality in Mainz;
and to M.\,Yu.~Kalmykov for bringing ref.~\cite{DK:01} to my attention
and discussions related to it.

\appendix
\section{Anticommuting $\gamma_5$ and 't~Hooft--Veltman $\gamma_5$}
\label{S:g5}

For flavour-nonsinglet currents one may use the anticommuting $\gamma_5$ without encountering contradictions;
they are related to the currents with the 't~Hooft--Veltman $\gamma_5$ by a finite renormalization~\cite{Larin}:
\begin{equation}
\left( \bar{q} \gamma_5^{\text{AC}} \Gamma_n \tau q \right)_\mu
= Z_{2-n}(\alpha^{(n_f)}(\mu))
\left( \bar{q} \gamma_5^{\text{HV}} \Gamma_n \tau q \right)_\mu\,,
\label{g5:Larin}
\end{equation}
where $\tau$ is a flavour matrix with $\Tr\tau=0$.
The currents with $\gamma_5^{\text{AC}} \Gamma_n$ have anomalous dimensions $\gamma_n$,
because they can be obtained from the case of massless quarks;
$\gamma_5^{\text{HV}} \Gamma_n$ is just $\Gamma_{4-n}$ with reshuffled components.
Equating the derivatives in $d\log\mu$ we obtain
\begin{equation}
Z_{2-n}(\alpha_s) = K_{\gamma_n-\gamma_{4-n}}^{(n_f)}(\alpha_s)\,,
\label{g5:Z}
\end{equation}
where the anomalous dimensions $\gamma_n$ and $\gamma_{4-n}$ differ starting from 2 loops.
In particular, $Z_0(\alpha_s)=1$.
In HQET currents  with $\gamma_5^{\text{AC}}$ and with $\gamma_5^{\text{HV}}$
have the same anomalous dimension $\tilde{\gamma}$,
and the finite renormalization factor similar to~(\ref{g5:Z}) is 1.
In the large $\beta_0$ limit (see~(\ref{b:gn}))
\begin{equation}
Z_n(\alpha_s) = \exp\left[ - \frac{8n}{\beta_0} \int_0^b db
\frac{(1 + \frac{2}{3} b) \Gamma(2+2b)}{(1+b)^2 (2+b) \Gamma^3(1+b) \Gamma(1-b)}
+ \mathcal{O}\left(\frac{1}{\beta_0^2}\right) \right]\,.
\label{g5:b}
\end{equation}

At the leading $1/\beta_0$ order we may use these formulae for flavour singlet currents, too.
The matrix $\gamma_5^{\text{AC}} \Gamma_n$ has the same property~(\ref{J:h}) but with $\eta=-(-1)^n$.
From our results~(\ref{b:f})--(\ref{b:N}) we see that, indeed,
\begin{equation}
\hat{H}_{\gamma_5^{\text{AC}} \Gamma_n} = \left.\hat{H}_{\Gamma_n}\right|_{\eta\to-\eta}
= \hat{H}_{\gamma_5^{\text{HV}} \Gamma_n} = \hat{H}_{\Gamma_{4-n}}\,.
\label{g5:r}
\end{equation}

Matrix elements of the currents with $\gamma_5^{\text{AC}}$ and $n=0$, 1 can be written via smaller numbers of form factors:
\begin{equation}
{<}Q(mv_2)|J|Q(mv_1){>} = F^P \bar{u}_2 \gamma_5^{\text{AC}} u_1\,,\quad
F^P = F_0 - 2 F_1 - (2w+1) F_2
\label{g5:P}
\end{equation}
(where $F_i$ with $n=0$, $\eta=-1$ are used), and
\begin{equation}
\begin{split}
&{<}Q(mv_2)|J^\mu|Q(mv_1){>} = F^A_1 \bar{u}_2 \gamma_5^{\text{AC}} \gamma^\mu u_1
+ F^A_2 \bar{u}_2 \gamma_5^{\text{AC}} u_1 \frac{(v_2 - v_1)^\mu}{2}\,,\\
&F^A_1 = F_0 + 2 F_1 + (2w-1) F_2\,,\quad
F^A_2 = 4 (F_1 - F_2)
\end{split}
\label{g5:A}
\end{equation}
(where $F_i$ with $n=1$, $\eta=1$ are used).

The divergence of the axial current is
\begin{equation}
i \partial_\mu \left(\bar{Q}_0 \gamma_5^{\text{AC}} \gamma^\mu Q_0\right)
= 2 m_0 \bar{Q}_0 \gamma_5^{\text{AC}} Q_0\,,
\label{g5:div}
\end{equation}
where the bare mass $m_0 = Z_m^{\text{os}} m$.
Taking the matrix element of this equation we obtain
\begin{equation}
F^A_1 + \frac{w-1}{2} F^A_2 = Z_m^{\text{os}} F^P\,.
\label{g5:me}
\end{equation}
The on-shell mass renormalization constant $Z_m^{\text{os}}$
at the first $1/\beta_0$ order
is given by the formula similar to~(\ref{b:f}), (\ref{b:fN})
with $N_m(\varepsilon,u) = - 2 (3-2\varepsilon) (1-u)$,
see, e.\,g., \cite{G:04}.
And indeed, from~(\ref{b:N}), (\ref{g5:P}--\ref{g5:A}) we obtain
\begin{equation}
N^A_1 + \frac{w-1}{2} N^A_2 = N^P + N_m\,.
\label{g5:N}
\end{equation}

\section{Expansion of the hypergeometric function $F$}
\label{S:F}

We can also find several terms of this expansion
using the \texttt{Mathematica} package \texttt{HypExp}~\cite{HypExp}
(which uses \texttt{HPL}~\cite{HPL}).
This results in
\begin{align}
&F = \frac{1}{\sinh\vartheta} \biggl[\biggr. \vartheta
- H_{-+}(\tau) u
- \left( H_{-+-}(\tau) - 2 H_{-+}(\tau) l \right) \frac{u^2}{2}
\nonumber\\
&{} - \left( H_{-+--}(\tau) - 2 H_{-+-}(\tau) l + 2 H_{-+}(\tau) l^2 \right) \frac{u^3}{3}
\nonumber\displaybreak\\
&{} - \left( H_{-+---}(\tau) - 2 H_{-+--}(\tau) l + 2 H_{-+-}(\tau) l^2 - \frac{4}{3} H_{-+}(\tau) l^3 \right) \frac{u^4}{4}
\nonumber\\
&{} - \biggl( H_{-+----}(\tau) - 2 H_{-+---}(\tau) l + 2 H_{-+--}(\tau) l^2 - \frac{4}{3} H_{-+-}(\tau) l^3
+ \frac{2}{3} H_{-+}(\tau) l^4 \biggr) \frac{u^5}{5}
\nonumber\\
&{} - \biggl( H_{-+-----}(\tau) - 2 H_{-+----}(\tau) l + 2 H_{-+---}(\tau) l^2 - \frac{4}{3} H_{-+--}(\tau) l^3 + \frac{2}{3} H_{-+-}(\tau) l^4\biggr.
\nonumber\\
&\qquad\biggl.{} - \frac{4}{15} H_{-+}(\tau)  l^5 \biggr) \frac{u^6}{6}
- \cdots \biggl.\biggr]\,,
\label{b:Fu}
\end{align}
where
\begin{equation}
\tau = \tanh\frac{\vartheta}{2}\,,\quad
l = \frac{1}{2} H_-(\tau) = \log\cosh\frac{\vartheta}{2}\,,\quad
H_+(\tau) = \vartheta\,,
\label{b:tau}
\end{equation}
and $H_{\cdots}(\tau)$ are harmonic polylogarithms (see~\cite{RV:00,HPL}).
Only one new polylogarithm appears at each order.

In order to compare the expansion coefficients in~(\ref{b:DK}) and in~(\ref{b:Fu}),
we need to transform them to harmonic polylogarithms of the same argument,
which we choose as $x = e^{-\vartheta}$.
In~(\ref{b:DK}), we first rewrite $S_{nm}(-x^{-1})$ via $S_{nm}(-x)$
using the formula from~\cite{DD:84};
then rewrite $S_{nm}(-x)$ via $H_{\cdots}(-x)$ and then via $H_{\cdots}(x)$;
rewrite $\log\cosh(\vartheta/2)$~(\ref{b:tau}) via $H_{\cdots}(x)$;
and finally re-express products of harmonic polylogarithms via their linear combinations.
In~(\ref{b:Fu}) we rewrite harmonic polylogarithms with $\pm$ indices~\cite{HPL}
via normal ones with indices 0, $\pm1$;
substitute $\tau=(1-x)/(1+x)$ and re-express via $H_{\cdots}(x)$;
and finally convert products of harmonic polylogarithms to sums.
All these steps are done in  \texttt{Mathematica} using \texttt{HPL}~\cite{HPL}.
We have checked that all the coefficients presented in~(\ref{b:Fu})
agree with~(\ref{b:DK}).

\section{Vector form factors}
\label{S:V}

The vector form factors $F^V_{1,2}$~(\ref{J:V}) can be written in the form~(\ref{b:f}), (\ref{b:fN});
from~(\ref{b:N}), (\ref{J:V}) we obtain
\begin{align}
N^V_1(\varepsilon,u) ={}& 2 \left[ 2 w + u - 3 u^2 - 3 w \varepsilon + 2 w u \varepsilon - (w-3) u^2 \varepsilon
+ w \varepsilon^2 - (w+1) u \varepsilon^2 \right] F
\nonumber\\
&{} - 2 \left[ 2 + u - 3 u^2 - 3 \varepsilon + 2 u \varepsilon + 2 u^2 \varepsilon
+ \varepsilon^2 - 2 u \varepsilon^2 \right]\,,
\label{V:N1}\\
N^V_2(\varepsilon,u) ={}& 4 u (1 + u - 2 u \varepsilon) F\,.
\label{V:N2}
\end{align}
All loop corrections to $F^V_1$ vanish at $\vartheta=0$,
and hence $N^V_1=0$ at $w=1$.

The form factor $F^V_1 = H^V_1/\tilde{Z}$,
where $\tilde{Z}$ at the $1/\beta_0$ order is determined by the anomalous dimension~(\ref{b:gh}),
and $H^V_1$ contains only non-negative powers of $\varepsilon$.
We choose $\mu=\mu'=\mu_0=m$.
$H^V_1$ at $\varepsilon=0$ is given by the coefficients
$f_{n0}$ (which produce $K_{-\tilde{\gamma}}$~(\ref{J:K}))
and $f_{0n}$ (which produce $\hat{H}^V_1$~(\ref{b:hh}));
$\varepsilon^n$ terms ($n>0$) require all $f_{nm}$.
Writing the expansion~(\ref{b:Fu}) as
$F = f_0 - f_1 u - f_2 u^2/2 - f_3 u^3/3 - \cdots$
we obtain up to 4 loops
\begin{align}
&H^V_1 = 1 + C_F \frac{b}{\beta_0} \Biggl\{
- 2 w f_1 + (3 w + 1) f_0 - 4
  - \left( w f_2 + (3 w + 1) f_1 - \left( \frac{\pi^2}{6} + 8 \right) w f_0 + \frac{\pi^2}{6} + 8 \right) \varepsilon
\nonumber\\
&\quad{}
  - \left( \frac{2}{3} w f_3 + \frac{3 w + 1}{2} f_2 + \left( \frac{\pi^2}{6} + 8 \right) w f_1
    + \left( \frac{2}{3} \zeta_3 w - \frac{\pi^2}{4} w - \frac{\pi^2}{12} - 16 w \right) f_0 - \frac{2}{3} \zeta_3 + \frac{\pi^2}{3} + 16 \right) \varepsilon^2
\nonumber\\
&\quad{}
  - \Biggl( \frac{w}{2} f_4 + \left( w + \frac{1}{3} \right) f_3 + \left( \frac{\pi^2}{12} + 4 \right) w f_2
    - \left( \frac{2}{3} \zeta_3 w - \frac{\pi^2}{4} w - \frac{\pi^2}{12} - 16 w \right) f_1
\nonumber\\
&\qquad{}
    - \left( \frac{\pi^4}{80} w - \zeta_3 w - \frac{1}{3} \zeta_3 + \frac{2}{3} \pi^2 w + 32 w \right) f_0
      + \frac{\pi^4}{80} - \frac{4}{3} \zeta_3 + \frac{2}{3} \pi^2 + 32 \Biggr) \varepsilon^3
  + \cdots
\nonumber\displaybreak\\
&{}
- b \Biggl[
w f_2 + \left( \frac{19}{3} w + 1 \right) f_1 - \frac{1}{3} \left( 2 \pi^2 w + \frac{209}{6} w + \frac{1}{2} \right) f_0
      + \frac{2}{3} \left( \pi^2 + \frac{53}{3} \right)
\nonumber\\
&\quad{}
  + \Biggl( 2 w f_3 + \frac{1}{2} \left( \frac{47}{3} w + 3 \right) f_2 + \left( \frac{3}{2} \pi^2 w + \frac{281}{9} w + \frac{1}{3} \right) f_1
\nonumber\\
&\qquad{}
    - \left( 8 \zeta_3 w + \frac{131}{36} \pi^2 w + \frac{3}{4} \pi^2 + \frac{5813}{108} w - \frac{203}{36} \right) f_0
    + 8 \zeta_3 + \frac{79}{18} \pi^2 + \frac{1301}{27} \Biggr) \varepsilon
\nonumber\\
&\quad{}
  + \Biggl( \frac{7}{2} w f_4 + \frac{1}{3} \left( \frac{103}{3} w + 7 \right) f_3 + \frac{1}{3} \left( \frac{19}{4} \pi^2 w + \frac{317}{3} w + 1 \right) f_2
\nonumber\\
&\qquad{}
    + \frac{1}{3} \left( 46 \zeta_3 w + \frac{271}{12} \pi^2 w + \frac{19}{4} \pi^2 + \frac{6677}{18} w - \frac{203}{6} \right) f_1
\nonumber\\
&\qquad{}
    - \frac{1}{3} \left( \frac{199}{80} \pi^4 w + \frac{317}{3} \zeta_3 w + 23 \zeta_3 + \frac{1693}{36} \pi^2 w + \frac{5}{12} \pi^2
      + \frac{129389}{216} w - \frac{6563}{72} \right) f_0
\nonumber\\
&\qquad{}
  + \frac{1}{3} \left( \frac{199}{80} \pi^4 + \frac{386}{3} \zeta_3 + \frac{427}{9} \pi^2 + \frac{27425}{54} \right) \Biggr) \varepsilon^2
  + \cdots
\Biggr]
\nonumber\\
&{}
- b^2 \Biggl[
\frac{4}{3} w f_3 + \left( \frac{19}{3} w + 1 \right) f_2 + \frac{1}{3} \left( 4 \pi^2 w + \frac{203}{3} w + 1 \right) f_1
\nonumber\\
&\qquad{}
    - \frac{1}{3} \left( 28 \zeta_3 w + \frac{38}{3} \pi^2 w + 2 \pi^2 + \frac{4919}{54} w - \frac{139}{6} \right) f_0
      + \frac{1}{3} \left( 28 \zeta_3 + \frac{44}{3} \pi^2 + \frac{1834}{27} \right)
\nonumber\\
&\quad{}
  + \Biggl( 6 w f_4 + 4 \left( \frac{52}{9} w + 1 \right) f_3 + \frac{1}{2} \left( 7 \pi^2 w + \frac{1171}{9} w + \frac{5}{3} \right) f_2
\nonumber\\
&\qquad{}
    + \left( 44 \zeta_3 w + \frac{359}{18} \pi^2 w + \frac{7}{2} \pi^2 + \frac{5366}{27} w - \frac{310}{9} \right) f_1
\nonumber\\
&\qquad{}
    - \left( \frac{92}{45} \pi^4 w + \frac{1114}{9} \zeta_3 w + 22 \zeta_3 + \frac{4075}{108} \pi^2 w + \frac{17}{36} \pi^2
      + \frac{258445}{972} w - \frac{9473}{108} \right) f_0
\nonumber\\
&\qquad{}
    + \frac{1}{9} \left( \frac{92}{5} \pi^4 + 1312 \zeta_3 + \frac{2063}{6} \pi^2 + \frac{43297}{27} \right)
  \Biggr) \varepsilon
  + \cdots
\Biggr]
\nonumber\\
&{}
- b^3 \Biggl[
3 w f_4 + 2 \left( \frac{19}{3} w + 1 \right) f_3 + \left( 2 \pi^2 w + \frac{203}{6} w + \frac{1}{2} \right) f_2
\nonumber\\
&\qquad{}
  + \left( 24 \zeta_3 w + \frac{38}{3} \pi^2 w + 2 \pi^2 + \frac{4955}{54} w - \frac{139}{6} \right) f_1
\nonumber\\
&\qquad{}
  - \left( \frac{71}{60} \pi^4 w + \frac{233}{3} \zeta_3 w + 12 \zeta_3 + \frac{203}{9} \pi^2 w + \frac{\pi^2}{3}
    + \frac{34937}{324} w - \frac{6007}{108} \right) f_0
\nonumber\\
&\qquad{}
  + \frac{1}{3} \left( \frac{71}{20} \pi^4 + 269 \zeta_3 + \frac{206}{3} \pi^2 + \frac{4229}{27} \right)
  + \cdots
\Biggr]
+ \cdots
\Biggr\}\,.
\label{V:F1}
\end{align}

The form factor $F^V_2=H^V_2$ is finite at $\varepsilon=0$
(this requirement explains why $N^V_2$~(\ref{V:N2}) vanishes at $u=0$).
We obtain
\begin{align}
&F^V_2 = C_F \frac{b}{\beta_0} \Biggl\{
2 f_0
  - 2 (f_1 - 4 f_0) \varepsilon
  - \left( f_2 + 8 f_1 - \left( \frac{\pi^2}{6} + 16 \right) f_0 \right) \varepsilon^2
\nonumber\\
&\quad{}
  - \frac{2}{3} \left( f_3 + 6 f_2 + \left( \frac{\pi^2}{4} + 24 \right) f_1 + \left( \zeta_3 - \pi^2 - 48 \right) f_0 \right) \varepsilon^3
  + \cdots
\nonumber\displaybreak\\
&{}
- b \Biggl[
  2 f_1 - \frac{25}{3} f_0
  + \left( 3 f_2 + \frac{74}{3} f_1 - \frac{1}{2} \left( 3 \pi^2 + \frac{961}{9} \right) f_0 \right) \varepsilon
\nonumber\\
&\quad{}
  + \frac{1}{3} \left( 14 f_3 + 86 f_2 + \left( \frac{19}{2} \pi^2 + \frac{1105}{3} \right) f_1
    - \left( 46 \zeta_3 + \frac{233}{6} \pi^2 + \frac{23545}{36} \right) f_0 \right) \varepsilon^2
  + \cdots
\Biggr]
\nonumber\\
&{}
- b^2 \Biggl[
2 f_2 + \frac{50}{3} f_1 - \frac{1}{3} \left( 4 \pi^2 + \frac{317}{3} \right) f_0
\nonumber\\
&\quad{}
  + \left( 8 f_3 + \frac{149}{3} f_2 + \left( 7 \pi^2 + \frac{1912}{9} \right) f_1
    - \left( 44 \zeta_3 + \frac{521}{18} \pi^2 + \frac{18451}{54} \right) f_0 \right) \varepsilon
  + \cdots
\Biggr]
\nonumber\\
&{}
- b^3 \Biggl[
4 f_3 + 25 f_2 + \left( 4 \pi^2 + \frac{317}{3} \right) f_1 - \left( 24 \zeta_3 + \frac{50}{3} \pi^2 + \frac{8609}{54} \right) f_0
+ \cdots
\Biggr]
+ \cdots
\Biggr\}\,.
\label{V:F2}
\end{align}

Using \texttt{HPL}~\cite{HPL} we have successfully reproduced all $n_l^{L-1} \alpha_s^L$ terms
with $L=1$, 2, 3 in $F^V_{1,2}$ from~\cite{HSSS:17}.

\end{document}